\documentclass[a4paper,11pt]{article}
\usepackage{pos}
\usepackage{amsmath}
\usepackage{graphicx} 
\usepackage{float}
\usepackage{eucal}
\usepackage{wrapfig}
\usepackage{caption}
\usepackage{mathrsfs}
\usepackage{lineno}

\title{Search for high-energy neutrino emission from hard X-ray AGN with IceCube}
 \ShortTitle{Search for neutrinos from hard X-ray AGN}

\author{The IceCube Collaboration \\{\normalsize \normalfont(a complete list of authors can be found at the end of the proceedings)}}

\emailAdd{sreetama.goswami@icecube.wisc.edu}

\abstract{ The IceCube Neutrino Observatory has detected high-energy astrophysical neutrinos in the TeV-PeV range. These neutrinos have an isotropic distribution on the sky, and therefore, likely originate from extragalactic sources. Active Galactic Nuclei form a class of astronomical objects which are promising neutrino source candidates given their high electromagnetic luminosity and potential ability to accelerate cosmic rays up to energies greater than 10$^{16}$ eV. Interactions of these cosmic rays within the AGN environment are expected to produce both neutrinos and pionic gamma rays. Some hadronic models of AGN emission suggest that such gamma rays can in turn interact with the dense photon fields of AGN and cascade down to hard X-rays and MeV gamma rays. We present an update on the IceCube stacking analysis searching for high-energy neutrinos from hard X-ray sources sampled from the {\it Swift}-BAT AGN Spectroscopic Survey.

\vspace{4mm}
{\bfseries Corresponding authors:}
Sreetama Goswami$^{1*}$, George C. Privon$^{2}$, Marcos Santander$^{1}$\\
{$^{1}$ \itshape Department of Physics and Astronomy, University of Alabama, Tuscaloosa, AL 35487 0324, USA}\\
{$^{2}$ \itshape North American ALMA Science Center, National Radio Astronomy Observatory, Charlottesville, VA 22903-2475, USA}\\[4mm]
$^*$ Presenter

\FullConference{37$^{\rm{th}}$ International Cosmic Ray Conference (ICRC 2021)\\
		July 12th -- 23rd, 2021\\
		Online -- Berlin, Germany}

}

\begin{document}
\maketitle



\section{Introduction}\label{sec:intro}

IceCube has detected astrophysical neutrinos in the TeV-PeV range that show an isotropic angular distribution, and their dominant origin remains unidentified ~\cite{Aartsen:2013jdh}. This points to extragalactic sources as a possible origin of the very-high-energy (VHE) neutrinos. Active Galactic Nuclei (AGN) is one such promising candidate source of VHE neutrinos. These are amongst the brightest objects in the universe owing to their high intrinsic luminosity. According to the unified model of AGN ~\cite{Netzer_2015}, these sources have a central supermassive black hole surrounded by an accretion disk and are sometimes accompanied by relativistic jets. Blazars are the AGN with the orientation of the jet lying close to our line of sight.    
AGN are powerful engines and capable of accelerating particles to very high energies. When the cosmic rays which are accelerated in the AGN environment interact with the ambient photon or matter field, they produce charged and neutral pions. The charged and neutral pions eventually decay to produce neutrinos and gamma rays, respectively, with comparable energies ~\cite{Kelner:2006tc, Kelner:2008ke}. Since gamma-ray emission at GeV-TeV energies has been observed from AGN, these are also likely emitters of VHE neutrinos.  

An IceCube study from 2016 searched for neutrino point sources from the second {\it Fermi}-LAT AGN catalog (2LAC) ~\cite{Aartsen:2016lir, Huber:2019lrm}. No significant excess was observed but upper limits were obtained for a contribution to the diffuse flux of cosmic neutrinos from blazars. However, a neutrino detected by IceCube on September 22, 2017, was found to be in temporal and spatial coincidence with a blazar, TXS 0506+056, that was in a flaring state ~\cite{IceCube:2018dnn}. This was the first evidence of a correlation between any astrophysical source and a high-energy neutrino. An examination of archival  IceCube data found excess neutrinos from the direction of the same blazar between September 2014 and March 2015,  when the blazar was not in a flaring state ~\cite{IceCube:2018cha}. Another study conducted with 10 years of IceCube data, searched for point-like sources of neutrinos and found an excess of neutrinos from a Seyfert II galaxy, NGC 1068, at a significance level of 2.9$\sigma$ ~\cite{Aartsen:2019fau}. This provides further evidence in favor of AGN  as  possible neutrino sources.

\begin{figure}
    \centering
    \includegraphics[width = 0.48\textwidth]{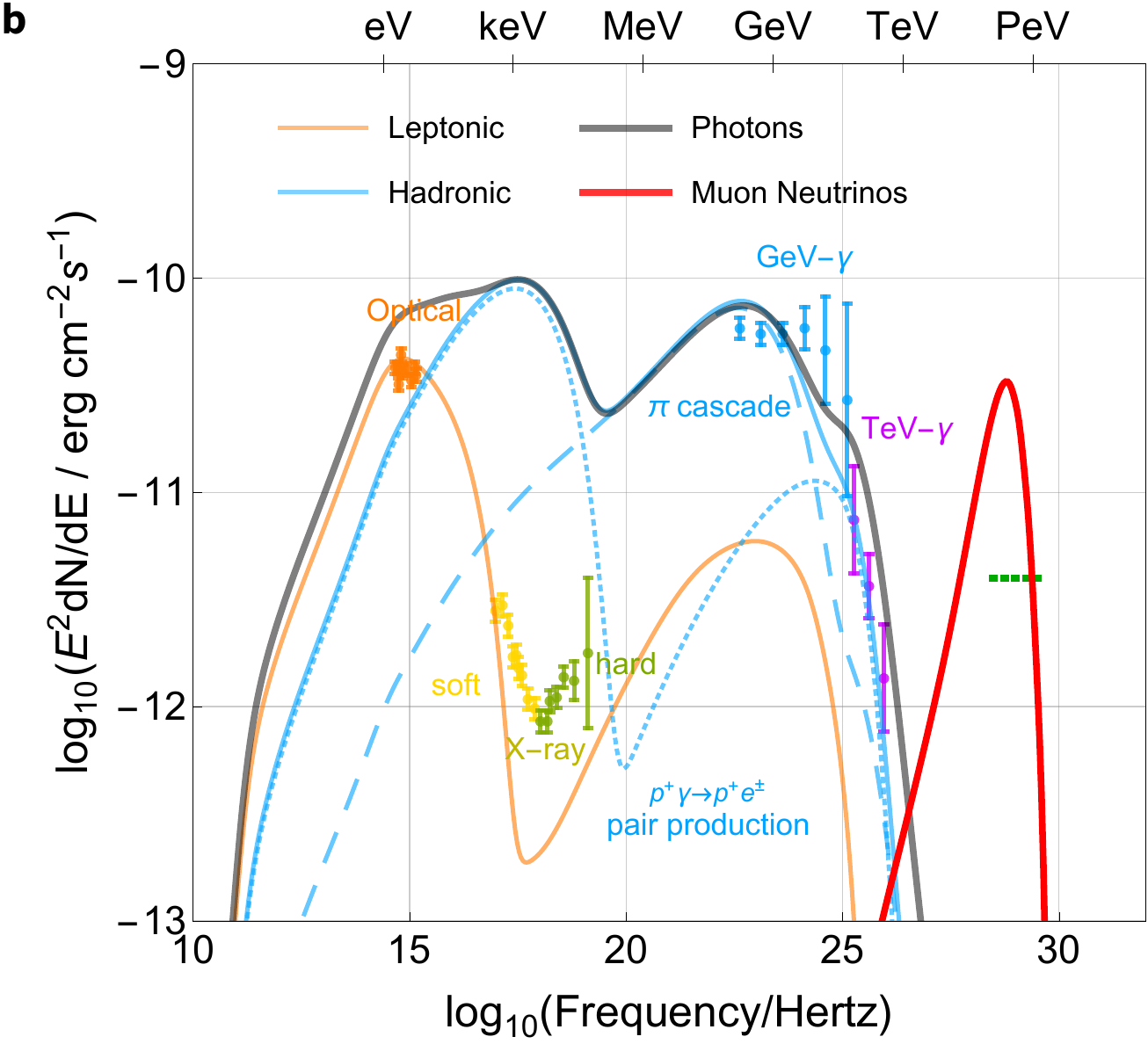}
    \caption{Spectral energy flux from TXS 0506+056 with different model fits. A simple  hadronic  model (in blue) used to fit the high energy peak  overestimates  the observed X-ray  flux. In this analysis, the focus is on sources selected in the 14-195 keV range (figure from \cite{Gao:2018mnu}). }
    \label{fig:sed_txs}
\end{figure}

There are various models that explain the production mechanism of high-energy neutrinos from AGN. Purely leptonic models can reproduce the spectral energy distribution (SED) of blazars  without neutrino production, while  purely hadronic models may fail to reproduce the SED completely (see Fig.\ref{fig:sed_txs}). 
According to some hybrid (leptohadronic) models, the TeV-PeV gamma rays produced  alongside neutrinos can interact with the ambient photons in AGN and cascade down to hard X-rays (high energy X-rays $\gtrsim$ 10 keV) or MeV gamma rays ~\cite{Petropoulou:2015upa, PetroMast2015, Murase:2018iyl}. This creates a scenario where we can no longer observe the TeV-PeV gamma rays but possibly find signatures of hard X-rays and MeV gamma rays. Hard X-rays could therefore be an important probe into the emission of neutrinos through hadronic processes.

\section{The catalog of hard X-ray sources}\label{sec:X-ray cat}

We select the sources for our analysis from the hard X-ray AGN detected by Burst Alert Telescope (BAT) onboard the Niel Gehrels {\it Swift} gamma-ray burst observatory ~\cite{Barthelmy:2005hs}. The {\it Swift}-BAT 70-Month hard X-ray Survey is the collection of objects observed by the BAT hard X-ray detector for a period of 70 months ~\cite{Baumgartner_2013}. There are 1171 sources detected in the > 10 keV band which are associated with 1210 ~counterparts. The counterparts were identified by an archival search for point sources (within $\sim$15 arcmin of BAT source position) in  X-ray observations of \textit{Swift}-XRT, \textit{Chandra}, \textit{XMM-Newton}, \textit{Suzaku} and ASCA. If no counterpart was found, then a 10 ks follow-up observation was carried out by \textit{Swift}-XRT using the BAT source coordinate. If the sources still did not have an observation, from either the archives or \textit{Swift}-XRT,  then NED and SIMBAD were checked to find a likely counterpart.

The BAT AGN Spectroscopic Survey\footnote{\href{https://www.bass-survey.com/}{https://www.bass-survey.com/}}, or the BASS catalog Data Release-1, is a selection of AGN sources from the 70-month BAT catalog ~\cite{Koss:2017ilk}. It has 838 AGN sources, mostly nearby, ~ 90$\%$ of them with a redshift below 0.2. Multi-wavelength physical properties of the sources have been determined by the analysis of the X-ray data along with the study of their available optical spectroscopic data ~\cite{Ricci_2017}.

\begin{figure}
    \centering
    \includegraphics[width=0.8\textwidth]{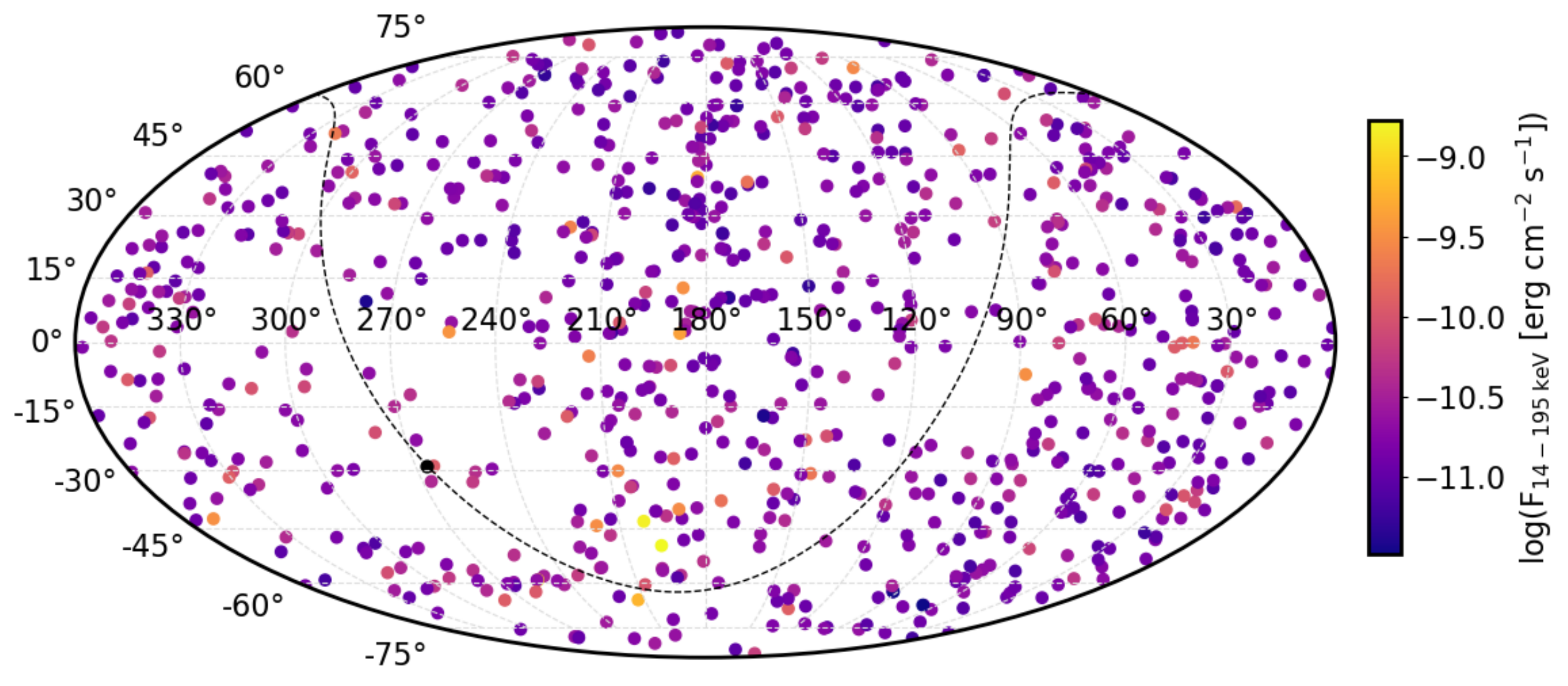} 
    \caption{Skymap in equatorial coordinates showing the BASS sources and their intrinsic flux in the hard X-ray band of energy 14-195 keV. The dashed black line shows the Galactic plane (figure from \cite{Santander:2019TE}).}
    \label{fig:skymap}
\end{figure}

The {\it Swift}-BAT survey presents a very uniform coverage of the sky. The BASS catalog is the most complete all-sky catalog of hard X-ray sources observed in the range of 14-195 keV and includes soft X-ray spectra in the range of 0.3-10 keV that are used to derive spectral parameters such as the column density.  For two  sources in the BASS catalog, there are no column density values since they do not have a soft X-ray observation in the 0.3-10 keV range carried out by {\it Swift-XRT, XMM-Newton-}EPIC/PN, or {\it Chandra}-ACIS. These sources are  SWIFT J1119.5+5132 and SWIFT J1313.6+3650A and they have not been included in our sample. Hence, we have 836 sources in our sample. A skymap of the BASS sources used in the study is shown in  Fig.\ref{fig:skymap}. 105 of the sources in the catalog are classified as blazars and 731 as non-blazar AGN. Of the blazars, 26 are BL Lac objects,  AGN whose optical spectrum is featureless, or with absorption lines that originate from the host galaxy and with weak and narrow emission lines; 53 are Flat Spectrum Radio Quasars (FSRQ), AGN with optical spectrum showing broad emission lines; and 26 blazars are of unidentified type.

\subsection{Possible overlap with other studies}\label{subsec:overlap}

\begin{figure}
\centering
\includegraphics[width=.6\linewidth]{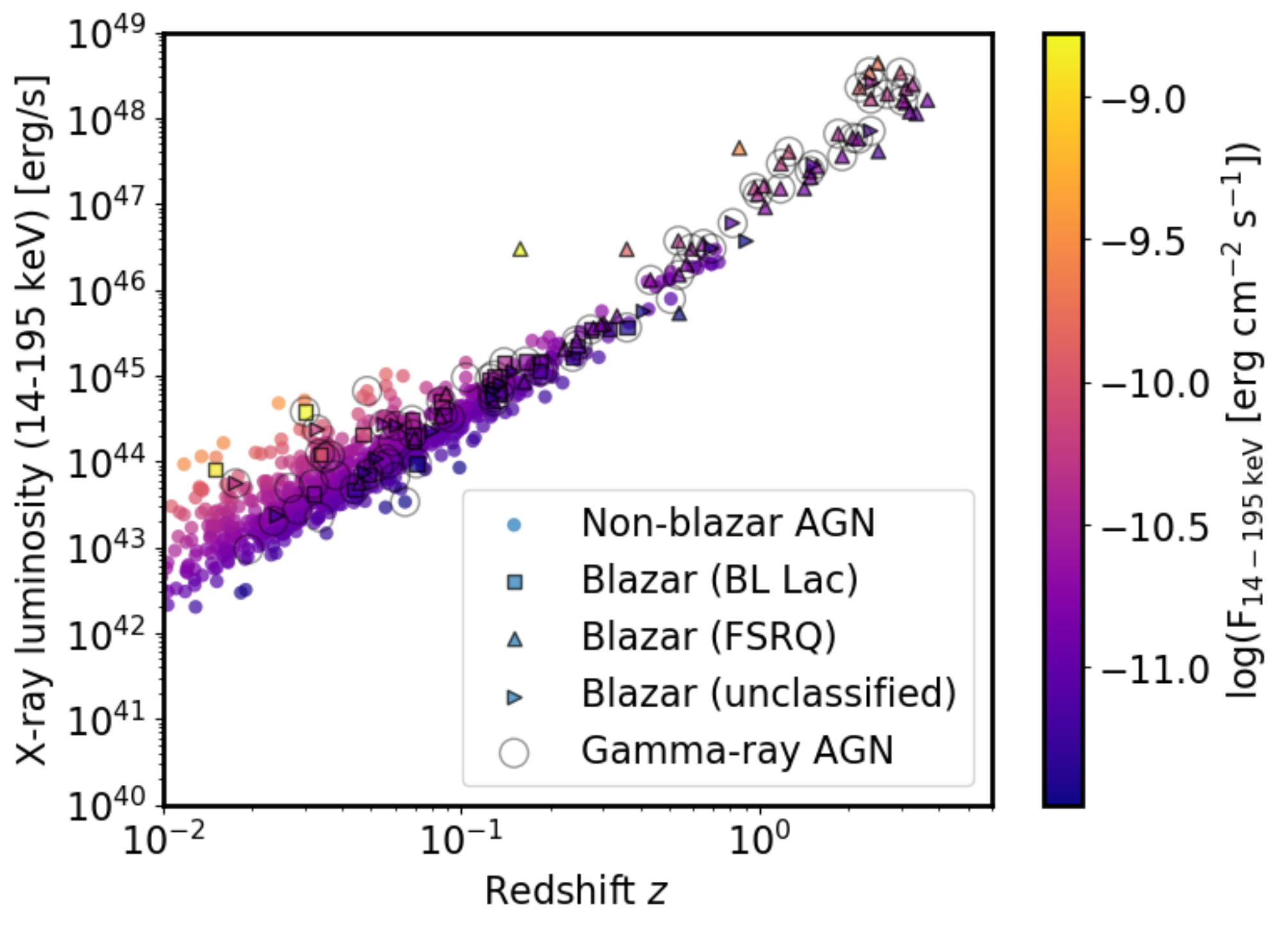} 
\caption{The figure shows the hard X-ray luminosity of the different classes of AGN in the BASS catalog as a function of their redshift.  The color represents the intrinsic flux of the sources in the energy range of 14-195 keV. AGN sources with likely counterparts in the \textit{Fermi}-LAT 4FGL catalog  are circled in gray (figure from \cite{Santander:2019TE}).}
\label{fig:lumi_z}
\end{figure}

AGN as a source class has been explored in an IceCube study that searched for neutrino emission from gamma-ray sources in the \textit{Fermi} 2LAC catalog and has found no significant correlation ~\cite{Aartsen:2016lir}. Several hard X-ray sources in the BASS catalog also exhibit gamma-ray emission (see Fig.\ref{fig:lumi_z}). We have evaluated the possible overlap with this previous IceCube study. 

A search was made for \textit{Fermi} sources in the 4FGL catalog ~\cite{Abdollahi_2020} located within a 0.2$^\circ$ circular region of interest (ROI) around the BASS sources. Since most of the gamma-ray sources in 4FGL have a lower angular uncertainty than 0.2$^\circ$, it is a conservative choice.  84 or 10$\%$ of the BASS sources have a 4FGL source within the ROI and 61 or 7$\%$ of the BASS sources are consistent with a 4FGL position considering its angular uncertainty. Most of the high-luminosity sources (most often blazars) are found in both catalogs. We conclude that the number of sources common with previous studies is not predominant. Thus, this work introduces a novel and unexplored class of neutrino source candidates.


\section{Analysis outline}\label{sec:ana_outline}

\subsection{IceCube dataset }

IceCube is a cubic-kilometer neutrino detector installed in the ice at the geographic South Pole ~\cite{Aartsen:2016nxy} between depths of 1450 m and 2450 m, completed in 2010. It contains 5160 digital optical modules (DOM) that detect Cherenkov radiation emitted by charged particles produced in the interactions of neutrinos in the surrounding ice or the nearby bedrock. This radiation is then used in the reconstruction of the direction, energy, and flavor of the neutrinos.

The diffuse flux of VHE neutrinos observed by IceCube includes neutrino-interaction events with different topologies such as cascades and tracks ~\cite{2013}. The tracks, ascribed to $\nu_{\mu}$ and anti-$\nu_{\mu}$, have typical angular resolutions of $<$ 1$^\circ$. The main source of background for the cosmic neutrino flux consists of atmospheric muons and atmospheric neutrinos, both produced in cosmic-ray interactions in the Earth's atmosphere ~\cite{PhysRevD.91.122004, Aartsen_2016}.

For this analysis, the IceCube dataset used is the gamma-ray follow-up (GFU) sample ~\cite{Aartsen_2017}. It solely contains  muon neutrino tracks with an all-sky coverage that have been reprocessed offline for better sensitivity. 7.5 years of the GFU  sample from the complete 86-string IceCube configuration between 2011-2018 have been used. 

\subsection{Likelihood approach}
The analysis involves a stacked search for point sources using a time-integrated unbinned maximum-likelihood method ~\cite{Achterberg_2006, Braun_2008}. The likelihood function is given by:
\begin{equation}
    \mathcal{L} (n_s, \gamma) = \prod_i^{N} \left(\frac{n_s}{N} \mathcal{S}_{i}(\vec{x_i},\sigma_i, E_i, \gamma) + \left(1-\frac{n_s}{N}\right) \mathcal{B}_{i}(\vec{x_i},E_i)\right).
\end{equation}

$\mathcal{S}$ and $\mathcal{B}$ represent the source and background probability density functions, respectively. Here, it is assumed that there is  $n_s$ number of signal neutrinos that follow a power-law spectrum $E\textsuperscript{-$\gamma$}$. The likelihood is evaluated using  data events with reconstructed direction and energy given by $\vec{x}_i$ and $E_i$, respectively, and $\sigma_i$ is the reconstructed directional uncertainty of each event. There is a total of $N$ neutrino events.

For a stacking analysis, the source PDF is a collective  contribution from all the sources combined into one term. Each source term  carries two kinds of weights; $ \omega ^{k}$, a theoretical weight depending on observed physical properties of the sources, and $R^{k}(\delta _{k},\gamma )$, a detector weight which is proportional to the effective area of the IceCube detector for a source at a declination $\delta $ with a spectral index $\gamma$. The stacked-source PDF for $M$ sources is given by:

\begin{equation}
    \mathcal{S}_{i,Stacked} =
      \frac{\sum_{k=1}^M \omega^k R^k(\delta_k,\gamma) \cdot
        \mathcal{S}_i^k(\vec{x}_i,\vec{x}_{k},\sigma_i,E_i, \gamma)}{\sum_{k=1}^M \omega^k R^k(\delta_k,\gamma)}.
\end{equation}

  Then, the likelihood $\mathcal{L}$ is maximized resulting in the best-fit values of the parameters,  $n_s=\hat{n}_s$ and $\gamma=\hat{\gamma}$. These values are used to calculate a Test Statistic ($\mathcal{TS}$) value according to the equation:
\begin{equation}
\mathcal{TS} = -2~\text{sgn}(\hat{n}_s) \log \left(\frac{\mathcal{L}(n_s=0)}
{\mathcal{L}(n_s=\hat{n}_s,\gamma=\hat{\gamma})} \right) .  
\end{equation}

The larger the value of the $\mathcal{TS}$, the less compatible is the data with a background-only hypothesis.

The background-only distribution of TS is evaluated by randomizing the coordinates of the data.   Only the right ascension, RA, of neutrino events are scrambled  to preserve the declination distribution of the events. Each scrambling process is a trial. To test the performance of the analysis, pseudo signals generated from Monte Carlo simulations following a power-law spectrum are injected into the data in the position of the BASS sources.\\
 
We plan to test different hypotheses with our analysis. Currently, we have only considered the hypothesis outlined below, and we present the other hypotheses in the outlook of the proceedings.

\noindent \textbf{Hypothesis I: }In this part, we use all  the 836 AGN from the BASS DR-1 catalog to perform the analysis. \textbf{Weighting Scheme:} 
    
\begin{itemize}
    \item[1.] {\bf  Flux weights}: This is proportional to the intrinsic flux in the hard X-ray range of 14-195 keV of the AGN sources. This is selective towards the sources with higher flux in this energy range.
    \item[2.] {\bf  Equal weights}: This is an unbiased weight that treats all sources in the catalog the same. 
\end{itemize}

\section{Current status: Sensitivity and Discovery Potential}

\begin{figure}
\centering
    \includegraphics[width = 0.48\textwidth]{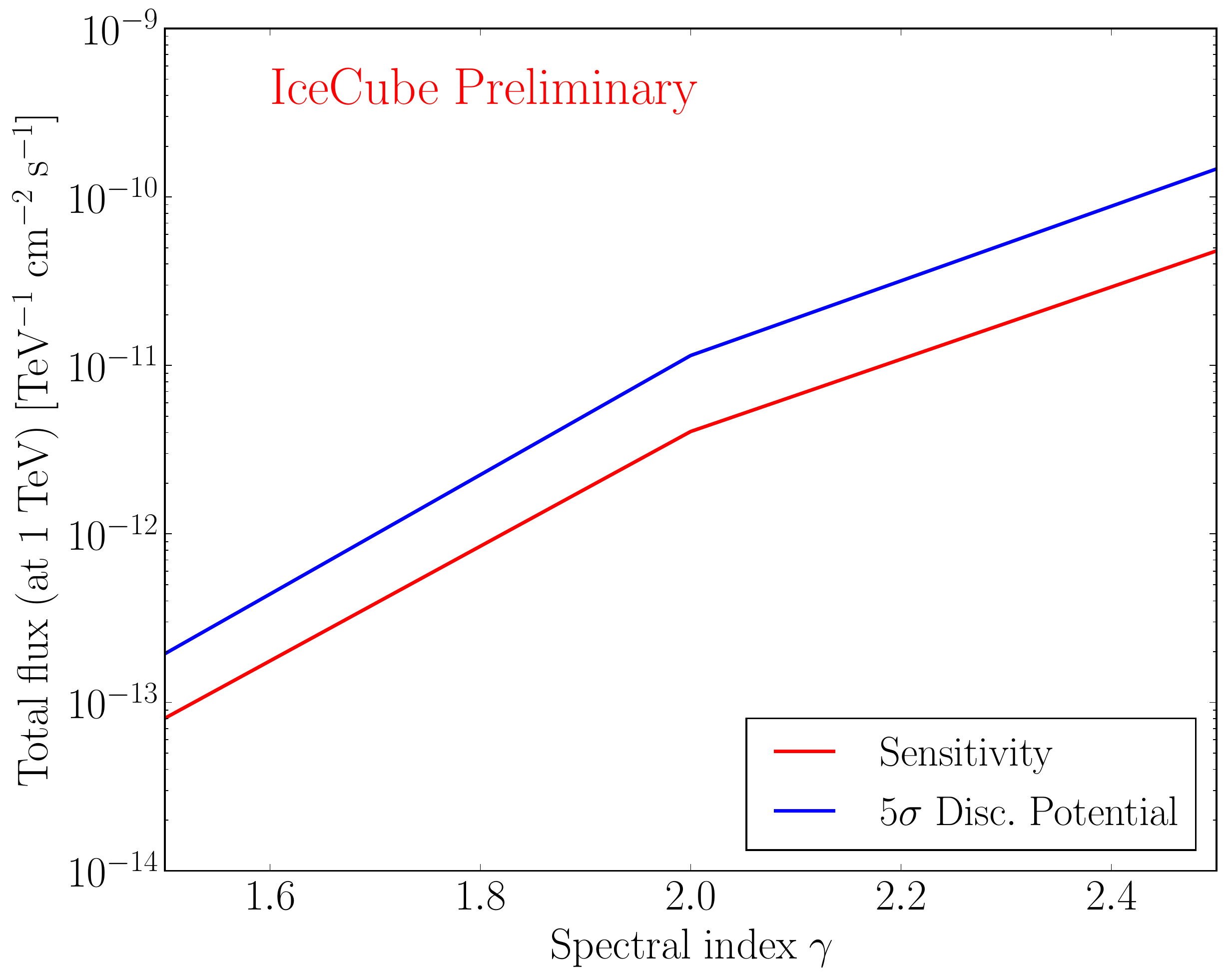}
    \includegraphics[width = 0.47\textwidth]{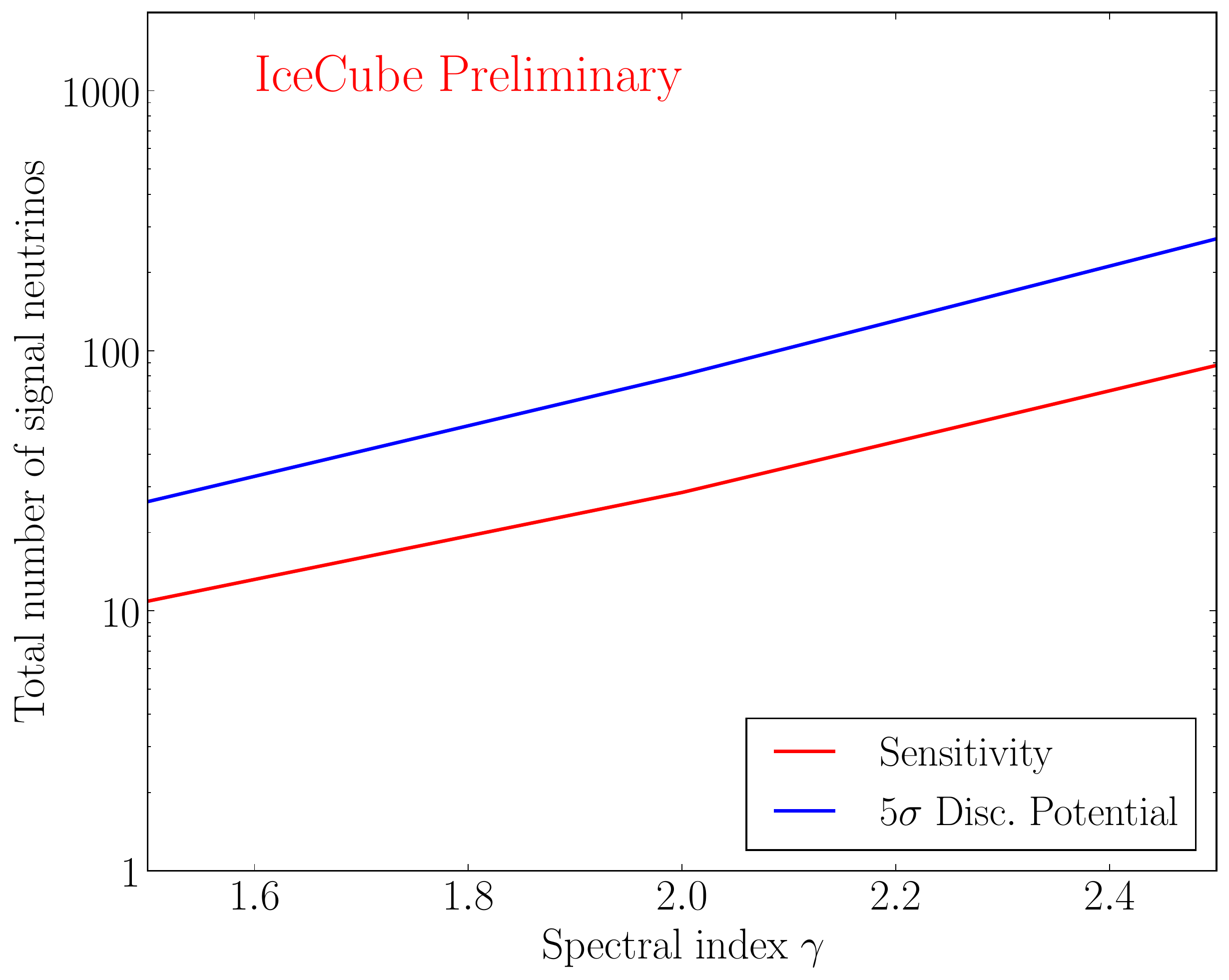}
    \caption{Plots showing the stacked flux at the normalization energy 1 TeV (left) and the number of signal neutrinos (right) as a function of the spectral index with the sensitivity and the 5$\sigma$  discovery potential  estimates, with Flux weights. }
    \label{fig:sens_disc_flux}
\end{figure}
\begin{figure}
\centering
    \includegraphics[width = 0.48\textwidth]{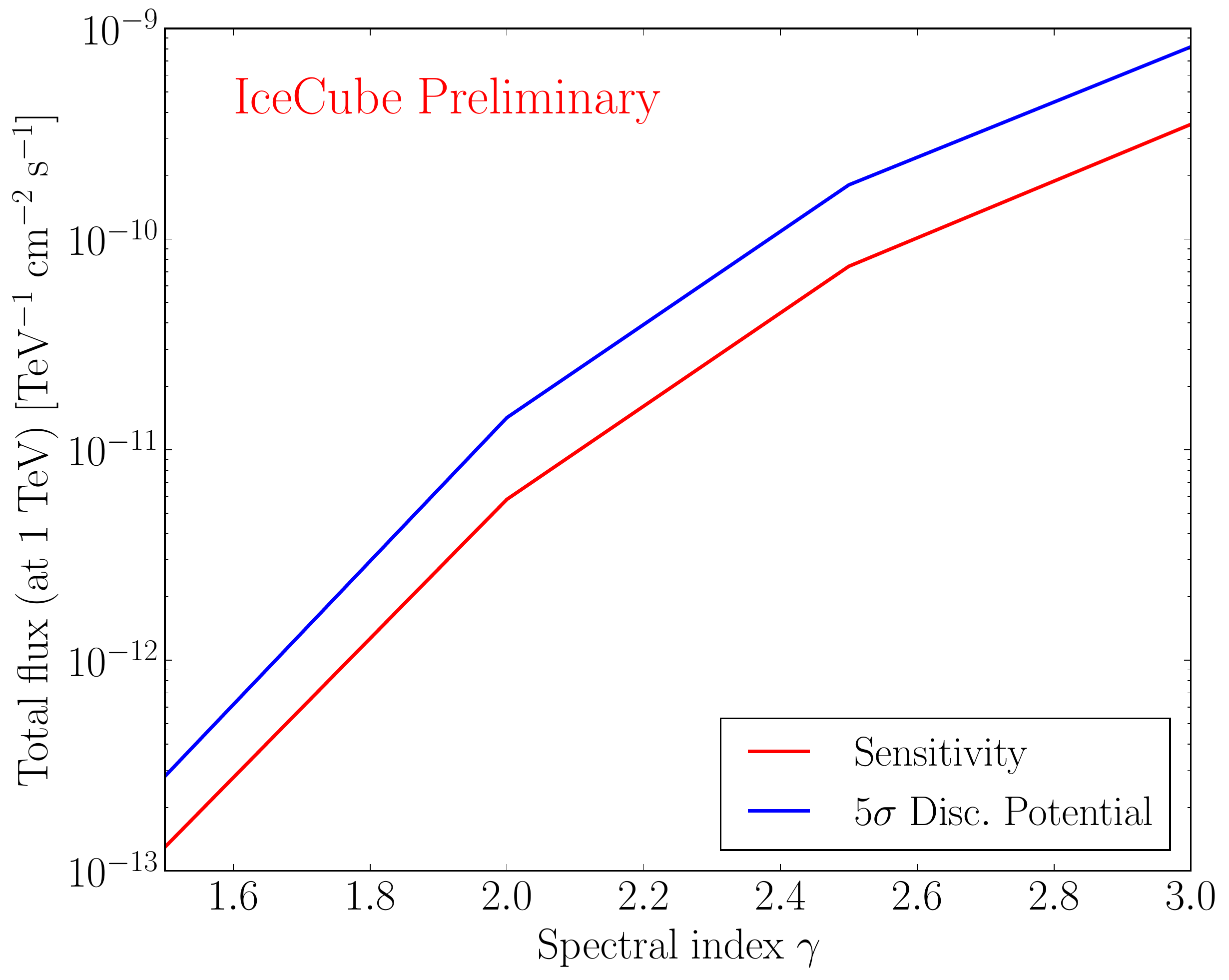}
    \includegraphics[width = 0.47\textwidth]{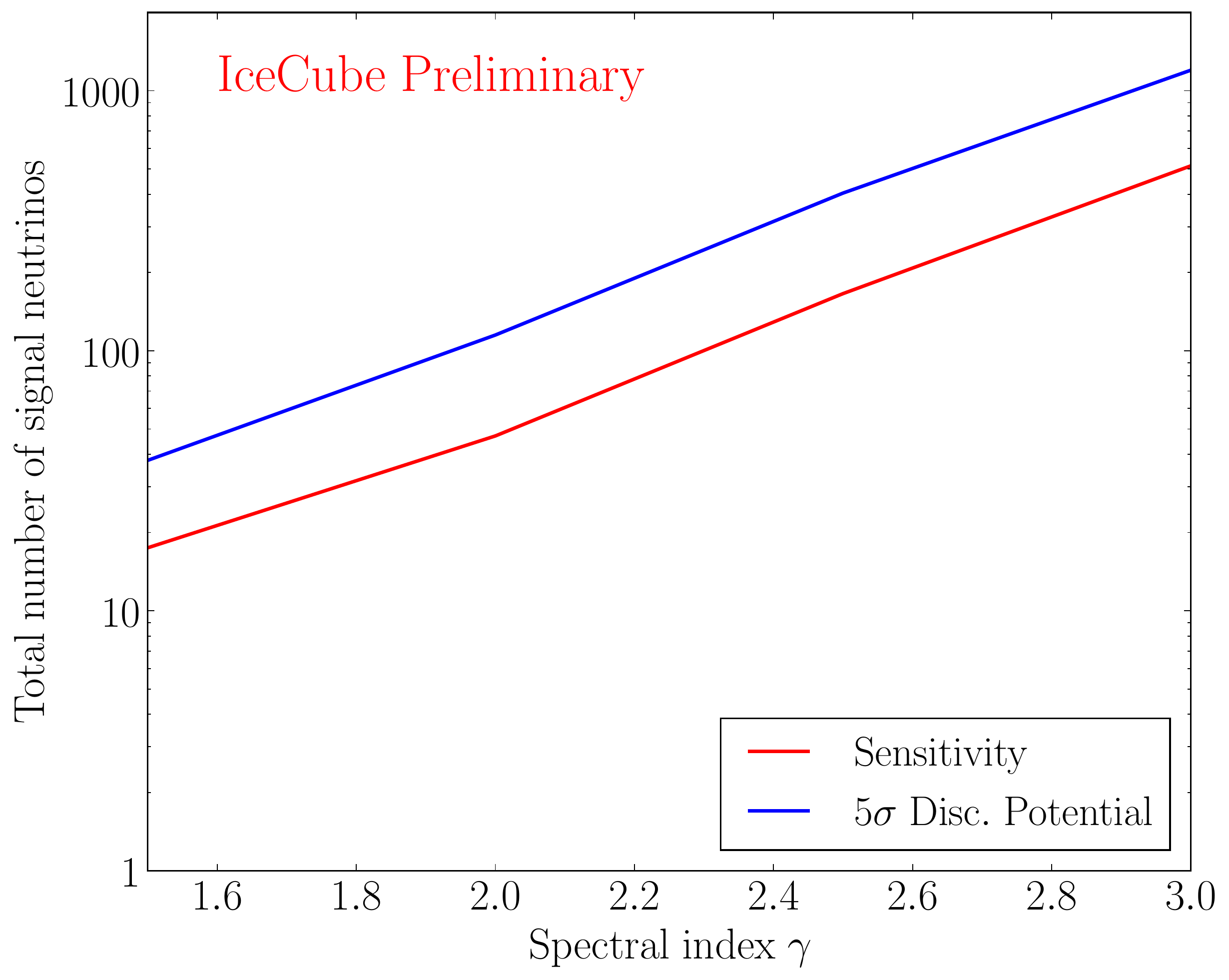}
    \caption{Plots showing the stacked flux at the normalization energy 1 TeV (left) and the number of signal neutrinos (right) as a function of the spectral index with the sensitivity  and the  5$\sigma$ discovery potential  estimates,  with Equal weights.}
    \label{fig:sens_disc_equal}
\end{figure}

The sensitivity and the $5\sigma$ discovery potential are  computed to give a sense of the power of the analysis performed. The sensitivity  is  defined as the flux required so that 90\% of injected signal trials have a value of TS higher than  in 50\% of the background-only trials. The $5\sigma$ discovery potential is the flux required so that 50\% of the time this flux will produce a result inconsistent with the background-only hypothesis at a significance of $5\sigma$.  Fig.\ref{fig:sens_disc_flux} and Fig.\ref{fig:sens_disc_equal} show the preliminary estimates for the sensitivity and the 5$\sigma$ discovery potential under Hypothesis I. 

\section{Outlook}

 In future work, we will perform  parts II and III of the analysis as described below.\vspace{0.2cm}
 
\noindent \textbf{Hypothesis  II:}  We will search for a correlation between the luminosity of the sources and their neutrino emission by performing  analyses on two of the following sub-classes of AGN in the catalog: Blazars, the AGN with high luminosity, and non-blazars, the AGN  with relatively lower luminosity. The categorization follows the classification presented in the {\it Swift}-BAT catalog. In this analysis, only {\bf Flux weights} will be applied and the results from each of the categories will be compared. \vspace{1em}

\noindent \textbf{Hypothesis III:} X-rays are an important tracer of AGN properties such as the neutral hydrogen column density (\textit{N}$_\text{H}$) which is a measure of the  obscuring material along the line of sight.  Obscuring matter can act as a target for the high-energy photons such as hard X-rays, partially attenuating and scattering these via Compton scattering and photoelectric absorption, and also affect the hadronic production of neutrinos. We will search for a correlation between the obscuring matter and the neutrino emission by dividing the sources in the catalog into three categories following the classification given in ~\cite{Ricci_2017}.  
Unobscured sources are the sources with a column density, $ N_{\text{H}} <$ 10$^{22}$ cm$^{-2}$; Obscured sources are the sources with a column density, 10$^{22}$ cm$^{-2}$ $\leq N_\text{H} <$  10$^{24}$ cm$^{-2}$; and Compton-thick sources are the sources with a column density,  $N_{\text{H}} \geq$ 10$^{24}$ cm$^{-2}$.
Results from the analysis of each of these categories will be compared. In this approach only {\bf Flux weights} will be applied.\\

From our analysis, we will either report the detection of a source and the associated neutrino flux or attain upper limits that constrain the contribution of AGN as a source class towards the diffuse flux of neutrinos observed by IceCube.



\small
\bibliographystyle{ICRC}
\bibliography{references}



\clearpage
\section*{Full Author List: IceCube Collaboration}




\scriptsize
\noindent
R. Abbasi$^{17}$,
M. Ackermann$^{59}$,
J. Adams$^{18}$,
J. A. Aguilar$^{12}$,
M. Ahlers$^{22}$,
M. Ahrens$^{50}$,
C. Alispach$^{28}$,
A. A. Alves Jr.$^{31}$,
N. M. Amin$^{42}$,
R. An$^{14}$
% ...
\\

\noindent
% ...
$^{12}$ Universit{\'e} Libre de Bruxelles, Science Faculty CP230, B-1050 Brussels, Belgium \\
% ...
$^{14}$ Department of Physics and Laboratory for Particle Physics and Cosmology, Harvard University, Cambridge, MA 02138, USA \\
% ...
$^{17}$ Department of Physics, Loyola University Chicago, Chicago, IL 60660, USA \\
$^{18}$ Dept. of Physics and Astronomy, University of Canterbury, Private Bag 4800, Christchurch, New Zealand \\
% ...
$^{22}$ Niels Bohr Institute, University of Copenhagen, DK-2100 Copenhagen, Denmark \\
% ...
$^{28}$ D{\'e}partement de physique nucl{\'e}aire et corpusculaire, Universit{\'e} de Gen{\`e}ve, CH-1211 Gen{\`e}ve, Switzerland \\
% ...
$^{31}$ Karlsruhe Institute of Technology, Institute for Astroparticle Physics, D-76021 Karlsruhe, Germany  \\
% ...
$^{42}$ Bartol Research Institute and Dept. of Physics and Astronomy, University of Delaware, Newark, DE 19716, USA \\
% ...
$^{50}$ Oskar Klein Centre and Dept. of Physics, Stockholm University, SE-10691 Stockholm, Sweden \\
% ...
$^{59}$ DESY, D-15738 Zeuthen, Germany \\
% ...

\subsection*{Acknowledgements}

\noindent
USA {\textendash} U.S. National Science Foundation-Office of Polar Programs,
U.S. National Science Foundation-Physics Division,
U.S. National Science Foundation-EPSCoR,
Wisconsin Alumni Research Foundation,
Center for High Throughput Computing (CHTC) at the University of Wisconsin{\textendash}Madison,
Open Science Grid (OSG),
Extreme Science and Engineering Discovery Environment (XSEDE),
Frontera computing project at the Texas Advanced Computing Center,
U.S. Department of Energy-National Energy Research Scientific Computing Center,
Particle astrophysics research computing center at the University of Maryland,
Institute for Cyber-Enabled Research at Michigan State University,
and Astroparticle physics computational facility at Marquette University;
Belgium {\textendash} Funds for Scientific Research (FRS-FNRS and FWO),
FWO Odysseus and Big Science programmes,
and Belgian Federal Science Policy Office (Belspo);
Germany {\textendash} Bundesministerium f{\"u}r Bildung und Forschung (BMBF),
Deutsche Forschungsgemeinschaft (DFG),
Helmholtz Alliance for Astroparticle Physics (HAP),
Initiative and Networking Fund of the Helmholtz Association,
Deutsches Elektronen Synchrotron (DESY),
and High Performance Computing cluster of the RWTH Aachen;
Sweden {\textendash} Swedish Research Council,
Swedish Polar Research Secretariat,
Swedish National Infrastructure for Computing (SNIC),
and Knut and Alice Wallenberg Foundation;
Australia {\textendash} Australian Research Council;
Canada {\textendash} Natural Sciences and Engineering Research Council of Canada,
Calcul Qu{\'e}bec, Compute Ontario, Canada Foundation for Innovation, WestGrid, and Compute Canada;
Denmark {\textendash} Villum Fonden and Carlsberg Foundation;
New Zealand {\textendash} Marsden Fund;
Japan {\textendash} Japan Society for Promotion of Science (JSPS)
and Institute for Global Prominent Research (IGPR) of Chiba University;
Korea {\textendash} National Research Foundation of Korea (NRF);
Switzerland {\textendash} Swiss National Science Foundation (SNSF);
United Kingdom {\textendash} Department of Physics, University of Oxford.

\end{document}